%% file: rtmc.tex
\documentclass{eptcs}
\usepackage{pst-all}
\usepackage{algorithm}
\usepackage{algorithmic}
\usepackage{amsmath}
\usepackage{macros}
\usepackage{amssymb}
\usepackage{verbatim}
\usepackage{url}
\usepackage{fixltx2e}

\title{A Formal Model For Real-Time Parallel Computation}

\author{
Peter Hui \qquad\qquad Satish Chikkagoudar
\institute{Pacific Northwest National Laboratory\\
Washington, USA}
\email{\quad peter.hui@pnnl.gov \quad\qquad satish.chikkagoudar@pnnl.gov}
}
\def\titlerunning{Formal Model for RT Parallel Computation}
\def\authorrunning{P. Hui \& S. Chikkagoudar}

\providecommand{\event}{FTSCS 2012}
\begin{document}
\maketitle
\input{abstract}

\input{background}

\input{formal}

\input{cases}

\input{remarks}

\bibliographystyle{eptcs}
\bibliography{bib}

\end{document}

%% file: abstract.tex
\begin{abstract}
The imposition of real-time constraints on a parallel computing environment--- 
specifically high-performance, cluster-computing systems---
introduces a variety of challenges with respect to the formal verification of the system's timing properties. 
In this paper, we briefly motivate the need for such a system, and we introduce an automaton-based method for performing such formal verification. We define the concept of a consistent parallel timing system: a hybrid system consisting of a set of timed automata (specifically, timed 
\buchi automata as well as a timed variant of standard finite automata), intended to model the timing properties of a 
well-behaved real-time parallel system. Finally, we give a brief case study to demonstrate the concepts in the paper: a 
parallel matrix multiplication kernel which operates within provable upper time bounds. 
We give the algorithm used, a corresponding consistent parallel timing system, and empirical results showing that the system operates under the specified timing constraints.
\end{abstract}

%% file: background.tex
\section{Introduction}
\label{sec:intro}
Real-time computing has traditionally been considered largely in the context of single-processor and embedded systems, and indeed, the terms real-time computing, embedded systems, and control systems are often mentioned in closely related
contexts. However, real-time computing in the context of multinode systems, specifically high-performance, cluster-computing systems, remains relatively unexplored. 
It can be argued that one reason for the relative dearth of work in this area 
is the lack of scenarios to date which would require such a system.
Previously \cite{rtss-wip, hui-sc}, we have motivated the emerging need for such an infrastructure, giving a specific
scenario related to the next generation North American electrical grid. 
In that work, we described the changes and challenges in the power grid driving the need
for much higher levels of \emph{computational resources} for power
grid operations.  To briefly summarize (and to provide some motivational
context for the current work), 
many of these computations--- particularly floating-point
intensive simulations and optimization calculations 
(\cite{fang,yousu, gao,gorton,ca})---
can be more
effectively done in a centralized manner, and 
the amount and scale of such data is 
estimated by some \cite{rtss-wip, hui-sc} to be on the order of terabytes per day 
of streaming sensor data (e.g. Phasor Measurement Units (PMUs)), with the need to analyze the data within a strict 
cyclical window (every 30ms), presumably with the aid of high-performance,
parallel computing infrastructures.
With this in mind, the current work is part of a larger research effort at
Pacific Northwest National Laboratory aimed at developing the necessary
infrastructure to support an HPC cluster environment capable of processing
vast amounts of streaming sensor data under hard real-time constraints. 

While verifying the timing properties of a more traditional (e.g. embedded) real-time system
poses complex questions in its own right,
imposing real-time constraints on a parallel (cluster) computing environment introduces 
an entirely new set of challenges not seen in these more traditional environments.
For example, in addition to standard real-time concepts such as 
\emph{worst-case execution time} (WCET), real-time parallel computation introduces the necessity of 
considering \emph{worst case transmission time} when communicating over the network between nodes,
as well as the need to ensure that timing properties of one process do not invalidate those of the
entire parallel process as a whole. 

These are but two examples of the many questions which must
be addressed in a real-time parallel computing system; certainly there are many more questions
than can be addressed in a single paper. To this end, we introduce a simple,
event driven, automata-based model of computation intended to model the timing properties of  
a specific class of parallel programs. Namely, we consider 
SPMD (Single Program, Multiple Data), parent-child type programs, in part because
in practice, many parallel programs--- including many prototypical MPI-based \cite{mvapich,mpi}
programs--- fall into this category. We give an example of such a program in Section~\ref{sec:cases}. 
This model is typified by the existence of a cyclic
\emph{master} or \emph{parent} process, and a set of noncyclic \emph{child} or \emph{slave}
processes amongst which work is divided. With this characterization, a very natural
correspondence emerges between the processes and the automata which model them: the cyclic \emph{parent} process
is very naturally modeled by an  $\omega$-automaton, and the \emph{child} processes by a 
standard finite automaton.  
Our main contribution of this paper, then, is twofold: first, a formal method of modeling the respective processes in this manner,
combining these into a single hybrid system of parallel automata, and secondly, a simple case
study demonstrating a practical application of this system.  We should note that the notion 
of parallel finite automata is not a new one; variants
have been studied before (e.g. \cite{paraut,pfa}).
We take the novel approach of combining \emph{timed} variants (\cite{tba,tioa}) of finite automata 
into a single hybrid model which captures the timing properties of the various
component processes of a parallel system.


The rest of the paper proceeds as follows: Section~\ref{sec:formal} defines the automaton models
used by our system: Timed Finite Automata in Section~\ref{sec:tafsa}, Timed \buchi Automata
in Section~\ref{sec:tba}, and a hybrid system combining these two models in Section~\ref{sec:pts}.
Section~\ref{sec:cases} gives a case study in the form of an example real-time matrix
multiplication kernel, running on a small, four-node real-time parallel cluster. 
Section~\ref{sec:concl} concludes.

%% file: formal.tex
\section{Formalisms}
\label{sec:formal}
In this section, we give formal definitions for the machinery used in our hybrid system of automata. The 
definitions given in Sections~\ref{sec:tafsa} and \ref{sec:tba} are not new \cite{tba}. However, it is still important
that we state their definitions here, as they are used later on, in Section~\ref{sec:pts}.

\input{tafsa}

\input{tba}

\input{tpba}

%% file: tafsa.tex
\subsection{Timed Finite Automata}
\label{sec:tafsa}
In this section, we define a simple timed extension 
of traditional finite state automata and the words they accept. We will use these 
in later sections to model the timing properties of child processes in a real-time 
cluster system.  

\emph{Timed strings} take the form $(\astring,\timeseq)$, where $\astring$ is a string of symbols, 
and $\timeseq$ is a monotonically increasing sequence of reals (timestamps).  
$\timing[x]$ denotes the timestamp at which symbol $\aalpha[x]$ occurs.  We 
also use the notation  $\pair{\aalpha[x]}{\timing[x]}$ to denote a particular symbol/timestamp pair.  For instance, the timed string $\pair{(abc)}{(1,10,11)}$ is equivalent to the sequence
$(a,1) (b,10) (c,11)$, and both represent the case where `a' occurs at time 1, `b' at time 10, 
and `c' at time 11.

Correspondingly, we extend traditional finite automata to include a set of 
\emph{timers}, which impose temporal restrictions along state transitions.
A timer can
be \emph{initialized} along a transition, setting its value to 0 when the transition is taken,
and it can be \emph{used} along a transition, indicating that the transition can only be taken
if the value of the timer satisfies the specified constraint. 
Formally, we associate with each automaton a set of timer variables $\atimers$, and 
following the nomenclature of \cite{tba}, an \emph{interpretation} $\clkint$ for this set
of timers is an assignment of a real value to each of the timers in $\clocks$. 
We write $\clkint\with{\map{\atimer}{0}}$ to denote the interpretation $\clkint$
with the value of timer $\atimer$ reset to 0.
Clock constraints consist of conjunctions of upper bounds:
\begin{definition}
For a set $\atimers$ of clock variables, the set $\constraints[\clocks]$ of clock constraints $\uconstraint$
is defined inductively as
\[
	\uconstraint := \LT{\atimer}{\aconst}\  |\  \uconstraint_1 \land \uconstraint_2
\]
where $\atimer$ is a clock in $\atimers$ and $\aconst$ is a constant in $\reals[+]$.
\end{definition}
While this definition may seem overly restrictive compared to
some other treatments (e.g. \cite{tba}), 
we believe it to be acceptable in this early work for
a couple of reasons. First, while simple, this sole syntactic form 
remains expressive enough to capture an interesting, non-trivial set of use cases (e.g.
Section~\ref{sec:cases}).
Secondly, the timing analysis 
in subsequent sections of the paper
becomes rather complex, even 
when timers are limited
to this single form. Restricting the syntax in this manner simplifies this analysis 
to a more manageable level. We leave more complex formulations and the corresponding 
analysis for future work.

\begin{definition}[Timed Finite Automaton (TFA)]
A Timed Finite Automaton (TFA) is a tuple 
\[\tuple{\Aalpha,\astates,\start,\accept,\clocks,\reln,\ireln,\creln}\], where 
\begin{itemize}
\item $\Aalpha$ is a finite alphabet, 
\item $\astates$ is a finite set of states,
\item $\start\in\astates$ is the start state,
\item $\accept\in\astates$ is the accepting state,
\item $\clocks$ is a set of clocks,
\item $\reln\subseteq \astates\times\astates\times \Aalpha$ is the state transition relation,
\item $\ireln \subseteq \reln\times\powset{\clocks}$ is the clock initialization relation, and
\item $\creln \subseteq \reln\times\constraints[\clocks]$ is the constraint relation.
\end{itemize}
A tuple $\tuple{\state[i],\state[j],\aalpha}\in\reln$ indicates that a symbol
$\aalpha$ yields a transition from state $\state[i]$ to state $\state[j]$, subject to the
restrictions specified by the timer constraints in $\creln$.
A tuple $\tuple{\state[i],\state[j],\aalpha, \set{\atimer[1],...,\atimer[n]}}\in\ireln$ indicates that on the
transition on symbol $\aalpha$ from $\state[i]$ to $\state[j]$, all of the specified timers
are to be initialized to $0$.
Finally, a tuple $\tuple{\state[i],\state[j],\aalpha, \constraints[\clocks]}\in\creln$ indicates that 
the transition on $\aalpha$ from $\state[i]$ to $\state[j]$ can only be
taken if the constraint $\constraints[\clocks]$ evaluates to true under the current timer interpretation.
\end{definition}

\begin{example}
The following TFA accepts the timed language 
$\setst{\pair{ab^*c}{\timing[1]...\timing[n]}}{\timing[n]-\timing[1] < 10}$
(i.e., the set of all strings consisting of an `a', followed by an arbitrary number of `b's,
followed by a `c', such that the elapsed time between the first and last symbols is no greater
than 10 time units).
The start state is denoted with a dashed circle, and the accepting state with a double line. 

\begin{center}
\begin{pspicture}(4,1.5)
\cnodeput[linestyle=dashed](0,0.4){n1}{$\state[1]$}
\cnodeput(2,0.4){n2}{$\state[2]$}
\cnodeput[doubleline=true](4,0.4){n3}{$\state[3]$}
\ncline{->}{n1}{n2}
\aput{:U}{a}
\bput{:U}{$\atimer=0$}

\ncline{->}{n2}{n3}
\aput{:U}{c}
\bput{:U}{$\timer{\atimer<10}$}

\ncarc[arcangleA=135,arcangleB=135,ncurv=5]{->}{n2}{n2}
\aput{:U}{b}
\end{pspicture}
\end{center}
\end{example}

Paths and runs are defined in the standard way:
\begin{definition}[Path]
Let $\atfa$ be a TFA with state set $\astates$ and transition relation $\reln$.
Then $\path{\state[1],...,\state[n]}$ is a \emph{path} over $\atfa$ if, for all 
$1 \le i < n$, $\some{\aalpha}{\tuple{\state_i, \state_{i+1},\aalpha} \in \reln}$.
\end{definition}

\begin{definition}[Run]
A run $\arun$ of a TFA
$\tfa{\Aalpha}{\astates}{\state[0]}{\accept}{\aclks}{\tbareln}{\ireln}{\creln}$
over a timed word $\pair{\astring}{\timeseq}$,
is a sequence of the form
\[
\arun: 
\pair{\state[0]}{\clkint[0]} 
  \runto{{\timing[1]}}{\aalpha[1]}
\pair{\state[1]}{\clkint[1]} 
  \runto{{\timing[2]}}{\aalpha[2]}
\pair{\state[2]}{\clkint[2]} 
  \runto{{\timing[3]}}{\aalpha[3]}
...
  \runto{{\timing[n]}}{\aalpha[n]}
\pair{\state[n]}{\clkint[n]} 
\]
satisfying the following requirements:
\begin{itemize}
\item Initialization: $\lookup{\clkint[0]}{\aclk}=0, \forall \aclk\in\clocks$
\item Consecution: For all $i\ge 0$: 
	\begin{itemize}
	\item $\tbareln\ni\tuple{\state[i],\state[i+1],\aalpha[i]}$, 
	\item $(\clkint[i-1] + \timing[i] - \timing[i-1])$ satisfies $\uconstraint_i$,
where $\creln\ni \tuple{\state[i-1],\state[i],\aalpha[i],\uconstraint_i}$, and 
	\item $\clkint[i] = (\clkint[i-1] + \timing[i] - \timing[i-1])\with{\map{\atimer}{0}}$, 
		$\forall \atimer \in \atimers$, where $\ireln\ni\tuple{\state[i],\state[j],\aalpha[i],\atimers}$
	\end{itemize}
\end{itemize}
$\arun$ is an \emph{accepting} run if $\state[n]=\accept$.
\label{def:runtfa}
\end{definition}

A TFA $\atfa$ \emph{accepts} a timed string 
$s=\pair{\astring}{\timing[1]...\timing[n]}$ if there is an accepting 
run of $s$ over $\atfa$, and $\timing_n-\timing_1$ is called the \emph{duration} of the string.

\begin{note}[Well-Formedness]
We introduce a restriction on how timers can be used in a TFA, thus defining 
what it means for a TFA to be \emph{well-formed}.
Namely, we restrict timers to be used only once along a path; this is to simplify somewhat
the timing analysis in subsequent sections.
In particular, we say that a TFA $\atfa$ is well-formed if, for all pairs of states 
$\pair{\state[x]}{\state[y]}$, all timers $\atimer$, and all paths from 
$\state[x]$ to $\state[y]$, $\atimer$ is used no more than once.
For example, the TFAs shown in Figure~\ref{fig:malformed} are not well-formed, since in both cases, timers can
potentially be used more than once--- in the first case ($\atfa[1]$), along the self-loop 
on $\state[2]$, and in the second case ($\atfa[2]$), along two separate transitions along the path.
At first, this may appear to be overly restrictive, but as it turns out, many of these
cases can easily be rewritten equivalently to conform to the single-use restriction, as
shown in Figure~\ref{fig:wellformed}.
\input{malformedTFA}

\input{wellformedTFA}

\end{note}

\input{delaybounds}

%% file: malformedTFA.tex
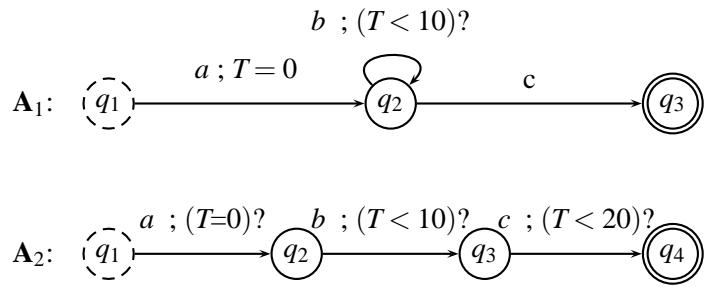
\begin{figure}[ht]
\begin{center}
\begin{pspicture}(0,0)(8.5,3.5)

\rput(0,2){$\atfa[1]$:}
\cnodeput[linestyle=dashed](1,2){q1}{$\state[1]$}
\cnodeput(4.75,2){q2}{$\state[2]$}
\cnodeput[doubleline=true](8.5,2){q3}{$\state[3]$}

\ncline{->}{q1}{q2}
\goto{\clock{a}{}{\atimer=0}}
\ncarc[arcangleA=135,arcangleB=135,ncurv=3]{->}{q2}{q2}
\goto{\clock{b}{\atimer < 10}{}}
\ncline{->}{q2}{q3}
\goto{c}

\rput(0,0){$\atfa[2]$:}
\cnodeput[linestyle=dashed](1,0){q4}{$\state[1]$}
\cnodeput(3.5,0){q5}{$\state[2]$}
\cnodeput(6,0){q6}{$\state[3]$}
\cnodeput[doubleline=true](8.5,0){q7}{$\state[4]$}
\ncline{->}{q4}{q5}
\goto{\clock{a}{$\atimer=0$}{}}
\ncline{->}{q5}{q6}
\goto{\clock{b}{\atimer<10}{}}
\ncline{->}{q6}{q7}
\goto{\clock{c}{\atimer<20}{}}

\end{pspicture}
\end{center}
\caption{Malformed TFAs.
Start states are denoted with a dashed circle, and accepting states with a double line.
The intent of $\atfa[1]$ is to allow
strings of the form $a$, followed by arbitrarily many $b$s, as long as they all 
occur less than 10 units after the $a$, followed by a $c$.  The intent of 
$\atfa[2]$ is to allows strings of the form $abc$, where the elapsed time between
the $a$ and $b$ is less than 10, and that between the $a$ and $c$ is less than 20.
Both of these can be rewritten using conforming automata, as shown in 
Figure~\ref{fig:wellformed}.}
\label{fig:malformed}
\end{figure}

%% file: wellformedTFA.tex
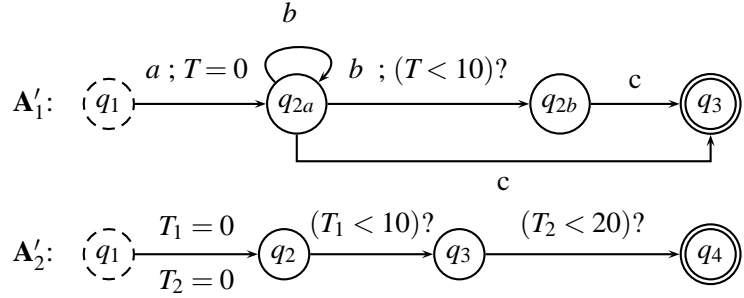
\begin{figure}[ht]
\begin{center}
\begin{pspicture}(0,0)(9,3)
\rput(0,2){$\atfa[1]'$:}
\cnodeput[linestyle=dashed](1,2){q1}{$\state[1]$}
\cnodeput(3.5,2){q2a}{$\state[2a]$}
\cnodeput(7,2){q2b}{$\state[2b]$}
\cnodeput[doubleline=true](9,2){q3}{$\state[3]$}

\ncline{->}{q1}{q2a}
\goto{\clock{a}{}{\atimer=0}}
\ncarc[arcangleA=135,arcangleB=135,ncurv=3]{->}{q2a}{q2a}
\goto{\clock{b}{}{}}
\ncline{->}{q2a}{q2b}
\goto{\clock{b}{\atimer<10}{}}
\ncline{->}{q2b}{q3}
\goto{c}

\ncangle[angleA=270,angleB=270]{->}{q2a}{q3}
\goto[c]{}

\rput(0,0){$\atfa[2]'$:}
\cnodeput[linestyle=dashed](1,0){q4}{$\state[1]$}
\cnodeput(3.33,0){q5}{$\state[2]$}
\cnodeput(5.66,0){q6}{$\state[3]$}
\cnodeput[doubleline=true](9,0){q7}{$\state[4]$}
\ncline{->}{q4}{q5}
\goto[{$\atimer[2]=0$}]{$\atimer[1]=0$}
\ncline{->}{q5}{q6}
\goto{\timer{\atimer[1]<10}}
\ncline{->}{q6}{q7}
\goto{\timer{\atimer[2]<20}}

\end{pspicture}
\end{center}
\caption{Equivalent, well-formed versions of automata from Figure~\ref{fig:malformed}.}
\label{fig:wellformed}
\end{figure}

%% file: delaybounds.tex
\subsubsection{Bounding Maximum Delay}
\label{sec:tfadelay}
An important notion throughout the remainder of the paper is that of
computing bounds on the allowable delays along all possible paths through a TFA. 
Specifically, we are interested in doing so to be able to reason formally about the maximum execution
time for a child process, with the end goal of being able to bound the execution time of the system--- parent and
all child processes--- as a whole. 

The idea is that we will ultimately use TFAs to represent the timing properties of a child
process. Paths through the automaton from its start state to an accepting state 
correspond to possible execution paths of the child process' code. Certainly, proving
a tight upper bound on the delay between two arbitrary points along an execution
path remains a very difficult problem, but to be clear, this is not our goal. Rather, our approach
involves modeling an execution path through a child process (and, by extension,
its corresponding timed automaton) using an \emph{event-based}
model, in which selected system events are modeled by transitions in the automaton,
and we rely on timing properties of the process to be guaranteed by the 
underlying RTOS process scheduler.
The problem of computing 
the worst-case delay through the automaton equates to
that of computing the maximum delay over all possible paths through the
automaton from its start state to its accepting state:
\[
\delayTFA[\atfa] = \max_{\apath\in\paths{\atfa}}
	{\pathdelay{\apath}{}{}}
\]
where 
\begin{itemize}
\item $\atfa=\tba{\Aalpha}{\astates}{\state[0]}{\state[f]}{\aclks}{\tbareln}{\ireln}{\creln}$ is the TFA
\item $\paths{\atfa}$ denotes the set of all paths in $\atfa$
from its start state $\state[0]$ to accepting state $\state[f]$, and 
\item $\pathdelay{\apath}{}{}$, for path $\apath=\path{\state[0],...,\state[f]}$, denotes the maximum delay 
from $\state[0]$ to $\state[f]$. That is, the maximum duration of any timed string $\pair{\astring}{\timeseq}$
such that $\pair{\state[0]...\state[f]}{\clkints}$ is a run of the string over $\atfa$ (for some $\clkints$).
\end{itemize}

\input{interactions}


%% file: interactions.tex
This problem can thus be formulated
in the following manner: given a timed finite automaton $\atfa$ and an 
integer $n$, is there a timed word of duration $d\ge n$ that is accepted by $\atfa$?
While simple cases, such as those presented in this paper, can be computed by observation and enumeration,
the complexity of the general problem
remains an open question, although we highly suspect it to be intractable--- Courcoubetis and Yannakakis give 
exponential-time algorithms for this and 
related problems, and have shown a strictly more difficult variant of the problem to be \PSPACE-complete 
\cite{minmax}. 
Furthermore, expanding the timer constraint syntax to a more 
expressive variant (c.f. \cite{tba}) can only complicate matters in terms of complexity.
We must be cautious, then, to ensure that we do not impose an inordinately large number of timers on a child process.

%% file: tba.tex
\subsection{Timed B\"{u}chi Automata}
\label{sec:tba}
Whereas we model the timing properties of the child processes of a cluster system using
the timed finite automata of the previous section, 
we model these properties of the parent using a timed variant of 
$\omega$-automata, specifically
Timed B\"uchi Automata. We assume a basic familiarity with these; due to space constraints, we give 
only brief overview here.  
To review briefly, 
$\omega$-automata, like standard finite automata, also consist of a finite number of states, 
but instead operate over words of infinite length. Classes of $\omega$-automata are distinguished
by their acceptance criteria. \buchi automata, which we consider in this paper, are defined
to accept their input if and only if a run over the input string visits an accepting state 
infinitely often. Other classes of $\omega$-automata exist as well. For example, Muller automata
are more stringent, specifying their acceptance criteria as a \emph{set} of acceptance sets; a
Muller automaton accepts its input if and only if the set of states visited infinitely often
is specified as an acceptance set.  More detailed specifics can be found elsewhere---
for example, \cite{tba}.

A \emph{Timed B\"uchi Automaton} (TBA) is a tuple 
$\tba{\Aalpha}{\astates}{\state[0]}{\accept}{\aclks}{\tbareln}{\ireln}{\creln}$, where
\begin{itemize}
\item $\Aalpha$ is a finite alphabet, 
\item $\astates$ is a finite set of states,
\item $\state[0]\in\astates$ is the start state,
\item $\accepts\subseteq\astates$ is a set of accepting states,
\item $\clocks$ is a set of clocks,
\item $\reln\subseteq \astates\times\astates\times \Aalpha$ is the state transition relation,
\item $\ireln \subseteq \reln\times\powset{\clocks}$ is the clock initialization relation, and
\item $\creln \subseteq \reln\times\constraints[\clocks]$ is the constraint relation.
\end{itemize}

A tuple $\tuple{\state[i],\state[j],\aalpha}\in\reln$ indicates that a symbol
$\aalpha$ yields a transition from state $\state[i]$ to state $\state[j]$, subject to the
restrictions specified by the clock constraints in $\creln$.
A tuple $\tuple{\state[i],\state[j],\aalpha, \clocks}\in\ireln$ indicates that on the
transition on symbol $\aalpha$ from $\state[i]$ to $\state[j]$, all clocks in 
$\clocks$ are to be initialized to $0$.
Finally, a tuple $\tuple{\state[i],\state[j],\aalpha, \constraints[\clocks]}\in\creln$ indicates that 
the transition on $\aalpha$ from $\state[i]$ to $\state[j]$ can only be
taken if the constraint $\constraints[\clocks]$ evaluates to true under the values of the current timer interpretation.

We define \emph{paths}, \emph{runs}, and \emph{subruns} over a TBA analagously to those over a TFA:
\begin{definition}[Path (TBA)]
Let $\atba$ be a TBA with state set $\astates$ and transition relation $\reln$.
$\path{\state[1],...,\state[n]}$ is a \emph{path} over $\atfa$ if, for all 
$1 \le i < n$, $\some{\aalpha}{\tuple{\state_i, \state_{i+1},\aalpha} \in \reln}$.
\end{definition}
\begin{definition}[Run, Subrun (TBA)]
A run (subrun) $\arun$, denoted by $\pair{\states}{\clkints}$, 
of a Timed B\"uchi Automaton
$\tba{\Aalpha}{\astates}{\state[0]}{\accept}{\aclks}{\tbareln}{\ireln}{\creln}$
over a timed word $\pair{\astring}{\timeseq}$,
is an infinite (finite) sequence of the form
\[
\arun: 
\pair{\state[0]}{\clkint[0]} 
  \runto{{\timing[1]}}{\aalpha[1]}
\pair{\state[1]}{\clkint[1]} 
  \runto{{\timing[2]}}{\aalpha[2]}
\pair{\state[2]}{\clkint[2]} 
  \runto{{\timing[3]}}{\aalpha[3]}
...
\]
satisfying the same requirements as given in Definition~\ref{def:runtfa}.
\label{def:runtba}
\end{definition}

For a run $\arun$, the set $\inf{\arun}$ denotes the set of states which are
visited infinitely many times. A TBA $\atba$ with final states $\accepts$
accepts a timed word $\aword=\pair{\astring}{\timeseq}$ if $\inf{\arun} \bigcap \accepts \neq \emptyset$, 
where $\arun$ is the run of $\aword$ on $\atba$. That is, a TBA accepts its input
if any of the states  from $\accepts$ repeat an infinite number of times in  $\arun$.

\input{tbaex1}

Lastly, we take the concept of maximum delay, introduced in the previous section with respect to
Timed Finite Automata, and extend it to apply to Timed B\"{u}chi Automata. Doing so first requires the following
definition, which allows us to restrict the timing analysis for TBAs to finite subwords: 
\begin{definition}[Subword over $\states$]
Let $\atba$ be a TBA, and let $\states=\path{\state[m]...\state[n]}$ be a finite path over $\atba$.
A finite timed word $\aword=\pair{(\aalpha[m]...\aalpha[n])}{(\timing[m]...\timing[n])}$ is a 
\emph{subword over} $\states$ iff 
$\exists{
	\state[0],...,\state[m-1],\aalpha[0],...,\aalpha[m-1], \timing[0],...,\timing[m-1]
}$
such that $\pair{
	\state[0]...\state[m-1]\state[m]...\state[n]
}{
	\clkints		
}$
is a subrun of $\pair{
	(\aalpha[0]...\aalpha[m-1]\aalpha[m]...\aalpha[n])
}{
	(\timing[0]...\timing[m-1]\timing[m]...\timing[n])
}$ over $\atba$ for some $\clkints$.
\label{def:subwordOver}
\end{definition}
Definition~\ref{def:subwordOver} is a technicality which is necessary to support the following 
definition of the maximum delay between states of a TBA:
\begin{definition}
Let $\atba$ be a TBA, and let $\states$ be a finite path over $\atba$.
Then $\delayTBA[\atba]{\states}$ is the maximum duration of any subword over 
$\states$.
\end{definition}
\begin{example}
Consider $\atba[1]$ from Example~\ref{ex:tba1}. Then
$\delayTBA[{\atba[1]}]{\state[1]\state[2]\state[2]\state[1]}=50$.
\end{example}

Algorithmically computing $\delayTBA[\atba]{\states}$ for a TBA $\atba$ is analogous to the case 
for TFAs;
in small cases (i.e., relatively few timers with small time constraints),
the analysis is relatively simple, while we conjecture the problem for more complex cases to be
intractable; we leave more detailed analysis for future work.

%% file: tbaex1.tex
\begin{example}
Consider the following TBA $\atba[1]$, with start state $\state[1]$ and accept states $F=\set{\state[1]}$: 
\begin{center}
\begin{pspicture}(0,-0.5)(5,2.5)
\cnodeput[doubleline=true](2,0){pq1}{$\state[1]$}
\cnodeput(5,0){pq2}{$\state[2]$}

\ncline{->}{pq1}{pq2}
\goto{\clock{a}{}{T=0}}

\ncarc[arcangleA=125, arcangleB=125, ncurv=3]{->}{pq2}{pq2}
\aput{:U}{\clock{b}{}{}}

\ncloop[angleB=180,loopsize=1.6]{->}{pq2}{pq1}
\dgoto{\clock{c}{T<50}{}}

\end{pspicture}
\end{center}
This TBA accepts the
$\omega$-language 
$L_1 = \setst{\pair{\infwd{(ab^*c)}}{\timing}}{
	\all{x}{
	  \some{i,j}{
	    \all{k}{
	      \boolform	
	    }
	  }
	}
}$
where $\boolform$ is the boolean formula 
\[\timing[i] < \timing[k] < \timing[j] \implies (\aalpha[i] = a) \land
				(\aalpha[k] = b) \land
				(\aalpha[j] = c) \land
				(\timing[j] -\timing[i] < 50)
\]

\label{ex:tba1}
\end{example}

%% file: tpba.tex
\subsection{Parallel Timing Systems}
\label{sec:pts}
Next, we model the timing properties of a SPMD-type parallel system as a whole by combining the two
models of Sections~\ref{sec:tafsa} and \ref{sec:tba} into a single \emph{parallel timing system}. 
A \emph{parallel timing system} (PTS) is a tuple $\tuple{\tbaparent,\atfas, \forkreln,\joinreln}$,
where 
\begin{itemize}
\item $\tbaparent=\tba{\Aalpha}{\astates}{\state[0]}{\accept}{\aclks}{\tbareln}{\ireln}{\creln}$ 
is a TBA (used to model the timing properties of the parent process)
\item $\atfas$ is a set $\set{\tbachild_1,...,\tbachild_n}$ of TFAs
(used to model the timing properties of the child processes)
\item $\forkreln\subseteq \reln\times{\atfas}$
 is a \emph{fork} relation (used to model the spawning of child processes)
\item $\joinreln\subseteq \reln\times{\atfas}$
 is a \emph{join} relation (used to model barriers (joins))
 \end{itemize}
A tuple $\tuple{\state[i],\state[j],\aalpha,\atfa}$ in $\forkreln$, with $\atfa\in\atfas$, indicates that an instance of
$\atfa$ is to be ``forked" on the transition 
from $\state[i]$ to $\state[j]$ on symbol $\aalpha$, and this ``fork" is denoted graphically as 
$\trans {\state[i]}{ \fork{\atfa} }{\state[j]}$,
modeling the spawning of a child process along the transition.
Similarly,
a tuple $\tuple{\state[i],\state[j],\aalpha,\atfa}$ in $\joinreln$ indicates that 
a previously forked instance of
$\atfa$ is to be ``joined" on the transition 
from $\state[i]$ to $\state[j]$ on symbol $\aalpha$.
This ``join" is denoted graphically as 
$\trans {\state[i]} { \join{\atfa} } {\state[j]}$,
modeling the joining along the transition with a previously spawned child process\footnote{
$\symfork$ was chosen as the symbol for `fork', as it graphically resembles a ``fork"; 
$\symjoin$ was chosen as that for `join', as it connotes ``ending" or ``finality".}.

\input{tpbaex1}

In theory, child processes could spawn children of their own (e.g. recursion).
For now, however, we disallow this possibility, as it somewhat complicates the analysis in 
the following section without adding significantly to the expressive power of the model.
The model can be expanded later to allow for arbitrarily nested children of children with the appropriate 
modifications; specifically, TBAs would need to be extended to include their own $\forkreln$ and $\joinreln$
relations, as would the definition of $\delayTFA$ for TBAs.

\input{tpba_consistency}

%% file: tpbaex1.tex
\begin{example}
Consider the following timing system $\apts[1]=\tuple{P,\set{\atfa},\forkreln,\joinreln}$:

\begin{pspicture}(0,-1)(14,2.5)

\rput(0,0){\rnode{P}{P:}}

\cnodeput[linestyle=dashed](2,0){pq1}{$\state[1]$}
\cnodeput[doubleline=true](5,0){pq2}{$\state[2]$}

\ncline{->}{pq1}{pq2}
\goto[\fork{$\atfa$}]{\clock{a}{}{T=0}}

\nccurve[angleB=120, angleA=60, ncurv=3]{<-}{pq2}{pq2}
\bput{:D}{\clock{b}{}{}}

\ncloop[angleB=180,loopsize=1.6]{->}{pq2}{pq1}
\dgoto[\join{$\atfa$}]{\clock{c}{T<50}{}}

\rput(7,0){\rnode{A}{$\atfa$:}}
\cnodeput[linestyle=dashed](8,0){aq1}{$\bstate[1]$}
\cnodeput(11,0){aq2}{$\bstate[2]$}
\cnodeput[doubleline=true](14,0){aq3}{$\bstate[3]$}

\ncline{->}{aq1}{aq2}
\goto{\clock{0}{}{U=0}}
\ncline{->}{aq2}{aq3}
\goto{\clock{1}{U<10}{}}
\end{pspicture}
\label{ex:system}

$P$ is the parent TBA with initial state $\state[1]$ and final state set $F=\{\state[2]\}$. 
$P$ accepts the $\omega$-language  $L_1$ (see p.~\pageref{ex:tba1}), and
$\atfa$ is a TFA which accepts the timed language $\setst{\pair{01}{\timing[1]\timing[2]}}{
	\timing[2]-\timing[1] < 10
}$.
In addition, the \emph{fork} and \emph{join} relations $\forkreln$ and $\joinreln$ dictate that on the transition from 
$\state[1]$ to $\state[2]$, an instance of $\atfa$ is forked ($\fork{\atfa}$), and that the transition from $\state[2]$
to $\state[1]$ can only proceed once that instance of $\atfa$ has completed ($\join{\atfa}$).

Conceptually, this system models a parent process ($P$) which exhibits periodic behavior,
accepting an infinite number of substrings of the form $ab^*c$, in which the initial `$a$'
triggers a child process $\atfa$ which must be completed prior to the end of the sequence, marked by
the following `$c$'. In addition, the `$c$' must occur no more than 50 time units after the initial `$a$'.
The child process is modeled by $\atfa$, which accepts strings of the form $01$, in which the $1$ must
occur no more than ten time units after the initial $0$. 
\label{ex:pts}
\end{example}

%% file: tpba_consistency.tex
Before proceeding, it is important to note that a PTS $\apts=\tuple{\tbaparent, \tbachildren, \forkreln,\joinreln}$ is not itself interpreted
as an automaton. In particular, we do not ever define a language accepted by $\apts$. Indeed,
it is not entirely clear what such a language would be, as we never specify the input to any of the children
in $\atfa$. Rather, the sole intent in specifying such a system $\apts$ is to specify the
\emph{timing behavior} of the overall system, rather than any particular language that would be accepted by it.

\subsubsection{Consistency}
With this said, we note that in Example~\ref{ex:pts}, $\atfa$ is in some sense 
``consistent" with its usage in $\tbaparent$. Specifically, since the maximum duration of any string accepted by 
$\atfa$ is 10, we are guaranteed that any instance of $\atfa$ forked on the 
$\trans{\state[1]}{a}{\state[2]}$ transition will have completed in time for the `join' along the 
$\trans{\state[2]}{c}{\state[1]}$ transition and hence, the timer $\timer{T<50}$ on this transition would be respected in all 
cases. In this sense, all $\pair{\fork{\atfa}}{\join{\atfa}}$ pairs are consistent
with timer $T$. However, such consistency is not always the case. Consider, for instance,
the parallel timing system $\apts[2]$ shown in Figure~\ref{fig:notcons}.
\input{tpbaex2}

In this case, there are two child processes: $\atfa$ and $\btfa$.
The maximum duration of a timed word accepted by $\atfa$ is 10, and that of 
$\btfa$ is 20. Supposing that an `a' occurs (and $\atfa$ forked) at time 0, it is thus
possible that the $\atfa$ will not complete until time $10-\epsilon_1$, at which time the `b' and fork of 
$\btfa$ can proceed. It is therefore possible that
$\btfa$ will not complete until time $30-\epsilon_1-\epsilon_2$ (for small $\epsilon_1,\epsilon_2$).
This would then violate the $\timer{\atimer<25}$ constraint,
corresponding to a case in which
a child process could take longer to complete than is allowable, given the timing constraints of the
parent process. It is precisely this type of interference
which we must disallow in order for a timing system to be considered consistent
with itself.  

To this end, we propose a method of defining \emph{consistency} within a timing system. 
Informally, we take the approach of 
deriving a new set of conditions from the timing constraints of the child processes,
so that checking \emph{consistency} reduces to the process of verifying that these
conditions respect the timing constraints of the master process.

First, we replace $\atfas, \forkreln,$ and $\joinreln$ from the  parallel timing system 
with a new set of \emph{derived} timers,
one for each $\atfa\in\atfas$, defining the possible ``worst case'' behavior of the child processes. 
Each such timer $\atimer[\atfa]$ is initialized on the transition along which
the corresponding
$\atfa$ is forked, and is used along (constrains) any transitions along which $\atfa$ is joined.
Each such use ensures that the timer is less than $\delayTFA[\atfa]$, representing the fact that
the elapsed time between the forking and joining of a child process is bounded in the worst case
by $\delayTFA[\atfa]$--- the longest possible duration for the child process. 
As an example, ``flattening" the 
timing system $\apts[1]$ of Example~\ref{ex:pts} results in a single new timer $\atimer[\atfa]$,
initialized along the $\trans{\state[1]}{a}{\state[2]}$ transition, and used along the
$\trans{\state[2]}{c}{\state[1]}$ transition with the constraint $\timer{\atimer[\atfa]<10}$.
We then check that none of these new derived timers invalidate the timing constraints of the 
parent process.

	Formally, we define two relations. The first of these is  \emph{flattening}, which takes 
	a parallel timing system $\pts{\tbaparent}{\tbachildren}{\forkreln}{\joinreln}$ 
	and yields a new pair of relations $\pair{\ireln}{\creln}$.
	Intuitively, $\ireln$ defines the edges along which each of the derived timers
	are initialized, and $\creln$ defines the edges along which each of the derived timers
	are used: 
\begin{definition}
Let  $\apts=\pts{\tbaparent}{\tbachildren}{\forkreln}{\joinreln}$ be a parallel timing system.
Then $\flatten{\apts} = \pair{\ireln}{\creln}$, where
\begin{align*}
\ireln&=  \setst{
		\tuple{\state[i],\state[j],\aalpha,\set{\atimer[\atfa]}}
	}{
		\tuple{\state[i],\state[j],\aalpha,\atfa} \in\forkreln
	} \\
\creln&= \setst{
		\tuple{\state[i],\state[j],\aalpha,\Constraint}
	}{
		\tuple{\state[i],\state[j],\aalpha,\atfa} \in\joinreln
	} 
\end{align*}
and
\begin{align*}
\Constraint &= 
	\bigwedge_{
					\tuple{\state[i],\state[j],\aalpha,\atfa} \in \joinreln
				}{
					(\atimer[\atfa] < \delayTFA[\atfa])
				}
\end{align*}
\end{definition}
The second relation takes $\forkreln$ and $\joinreln$ as inputs and extracts a set of edge pairs, defined
such that each such pair
$\pair{\aedge[1]}{\aedge[2]}$ specifies when a derived timer is initialized $(\aedge[1])$
and used $(\aedge[2])$. 
\begin{definition}
Let  $\apts=\tuple{\tbaparent,\tbachildren,\forkreln,\joinreln}$ be a parallel timing system,
with $\atfa\in\tbachildren$.
Then the set of all use pairs of $\atfa$ in $\apts$ is defined as $\uses[\atfa]{\apts} = \setst{
	\pair{
		\pair{\state[x]}{\state[y]}
	}{
		\pair{\state[m]}{\state[n]}
	}
}{
	(\tuple{\state[x],\state[y],\aalpha[1],\atfa} \in \forkreln) \land
	(\tuple{\state[m],\state[n],\aalpha[2],\atfa} \in \joinreln) 
}$ for some $\aalpha[1],\aalpha[2]$. Furthermore,
\[
	\uses{\apts} = \bigcup_{
		\atfa\in\tbachildren
	}{
		\uses[\atfa]{\apts}
	}
\]
\end{definition}

\begin{figure*}[tlh]
\begin{center}
\begin{pspicture}(10.5,7)
\rput(0,4){\rnode{P}{\tbaparent:}}

\cnodeput[linestyle=dashed](1,4){pq1}{$\state[1]$}
\cnodeput[doubleline=true](5,4){pq2}{$\state[2]$}
\cnodeput(5,7){pq3}{$\state[3]$}
\ncline{->}{pq1}{pq2}
\goto[$\fork{\atfa}$]{\clock{a}{}{\atimer=0}}
\ncline{->}{pq2}{pq3}
\bgoto[$\fork{\btfa}$]{$b$}
\ncline{->}{pq3}{pq1}
\dgoto[$\join{\atfa,\btfa}$]{\clock{c}{\atimer<24}{}}

\rput(0,0.5){\rnode{A}{$\atfa$:}}
\cnodeput[linestyle=dashed](1,1){aq1}{$\bstate[1]$}
\cnodeput(3,1){aq2}{$\bstate[2]$}
\cnodeput(4,0){aq3}{$\bstate[3]$}
\cnodeput(6,0){aq4}{$\bstate[4]$}
\cnodeput(4,2){aq5}{$\bstate[5]$}
\cnodeput(8,2){aq6}{$\bstate[6]$}
\cnodeput[doubleline=true](10.5,1){aq7}{$\bstate[7]$}

\ncline{->}{aq1}{aq2}
\goto{\clock{0}{}{U=0}}
\ncline{->}{aq2}{aq3}
\goto[0]{}
\ncline{->}{aq3}{aq4}
\goto[0]{}
\ncline{->}{aq4}{aq7}
\goto[\clock{0}{\btimer<10}{}]{}
\ncline{->}{aq2}{aq5}
\goto{1}
\ncline{->}{aq5}{aq6}
\goto{\clock{1}{\btimer<20}{\ctimer=0}}
\ncline{->}{aq6}{aq7}
\goto{\clock{1}{\ctimer<5}{}}
\ncline{->}{aq5}{aq4}
\goto{\clock{0}{}{}}

\rput(6,4){\rnode{B}{$\btfa$:}}
\cnodeput[linestyle=dashed](7,5){bq1}{$\cstate[1]$}
\cnodeput(9,5){bq2}{$\cstate[2]$}
\cnodeput[doubleline=true](11,5){bq3}{$\cstate[3]$}

\ncline{->}{bq1}{bq2}
\goto{\clock{0}{}{\ctimer=0}}
\ncline{->}{bq2}{bq3}
\goto{\clock{0}{\ctimer<11}{}}
\end{pspicture}
\end{center}
\caption{Parallel timing system $\apts[3]$. $\delayTFA[\atfa]=25, \delayTFA[\btfa]=11$.}
\label{fig:s3}
\end{figure*}
\begin{example}
Consider parallel timing system $\apts[3]$ shown in Figure~\ref{fig:s3}. 
Observe that $\delayTFA[\atfa]=25$ and $\delayTFA[\btfa]=11$. Then:
$\flatten{\apts[3]}=\pair{\ireln}{\creln}$, where
\begin{align*}
\ireln&=\set{\tuple{\state[1],\state[2],a,\set{{\atimer[\atfa]}}},
	\tuple{\state[2],\state[3],b,\set{{\atimer[\btfa]}}}}\\
\creln&=\set{
	\tuple{
		\state[3],\state[1],c,\Constraint}
}\text{ , where } \Constraint=(\atimer[\atfa] < 25)\land(\atimer[\btfa] < 11) 
\end{align*}
shown graphically in Figure~\ref{fig:s3-flat}, and 
\begin{align*}
\uses{\apts} &= \uses[\atfa]{\apts} \cup \uses[\btfa]{\apts} \\
&= \set{
	\pair{
		\pair{\state[1]}{\state[2]}
	}{
		\pair{\state[3]}{\state[1]}
	}
} \cup \set{
	\pair{
			\pair{\state[2]}{\state[3]}
		}{
			\pair{\state[3]}{\state[1]}
		}
}\\
&= \set{ \pair{
		\pair{\state[1]}{\state[2]}
	}{
		\pair{\state[3]}{\state[1]}
	}, 
	\pair{
			\pair{\state[2]}{\state[3]}
		}{
			\pair{\state[3]}{\state[1]}
		}
}
\end{align*}

\end{example}

\begin{figure}[tlh]
\begin{center}
\begin{pspicture}(0,0)(5,3.5)
\cnodeput[linestyle=dashed](0,0.5){pq1}{$\state[1]$}
\cnodeput[doubleline=true](5,0.5){pq2}{$\state[2]$}
\cnodeput(5,3.5){pq3}{$\state[3]$}
\ncline{->}{pq1}{pq2}
\goto[{\atimer=0\timersep\atimer[\atfa]=0}]{\clock{a}{}{}}
\ncline{->}{pq2}{pq3}
\bgoto[{\atimer[\btfa]=0}]{b}
\ncline{->}{pq3}{pq1}
\dgoto[${\timer{(\atimer[\atfa]<25) \land (\atimer[\btfa]<11)}}$]{\clock{c}{\atimer<24}{}}
\end{pspicture}
\end{center}
\caption{The result of flattening $\apts[3]$: forks and joins of $\atfa$ and $\btfa$ are shown along with
derived timers $\atimer[\atfa]$ and $\atimer[\btfa]$. Compare with Figure~\ref{fig:s3}.}
\label{fig:s3-flat}
\end{figure}

We can now proceed with a formal definition of consistency for a parallel timing system. Recall that
intuitively, such a system is consistent if the worst case timing scenarios over all child processes
will not invalidate the timing constraints of the parent process--- in other words, if the maximum delay 
between two states allowed
by the child processes never exceeds the corresponding maximum delay allowed by the timers in the 
parent process.

\begin{definition}[Consistency]
Let $\apts=\tuple{\atba,\forkreln,\joinreln,\tbachildren}$ be a PTS, where
\begin{itemize}
\item $\atba= \tba{\Aalpha}{\astates}{\state[0]}{\accepts}{\aclks}{\tbareln}{\ireln}{\creln} $ is a TBA
\item $\flatten{\apts} = \pair{\ireln'}{\creln'}$
\item $\atba'= \tba{\Aalpha}{\astates}{\state[0]}{\accepts}{\aclks}{\tbareln}{\ireln'}{\creln'}$
\end{itemize}
%
%
Then $\apts$ is \emph{consistent} if 
for all edge pairs $\pair{
	\pair{\state[x]}{\state[y]}
}{
	\pair{\state[m]}{\state[n]}
}\in\uses{\apts}$ and all paths $\apath=\state[x]\state[y]...\state[m]\state[n]$ through $\atba$,
\begin{equation}
	\delayTBA[\atba']{\apath} \le \delayTBA[\atba]{\apath}
\label{eq:cons}	
\end{equation}
\label{def:cons}
\end{definition}

We conclude this section with a few simple examples, which should help to clarify 
Definition~\ref{def:cons}; the following section gives a more realistic example.
\input{thm_cons}

\input{thm_incons}

%% file: tpbaex2.tex
\begin{figure}[lht]
\begin{center}
\begin{pspicture}(0,1)(14,3)

\rput(-1,1){\rnode{P}{P:}}

\cnodeput[linestyle=dashed](0.25,1){pq1}{$\state[1]$}
\cnodeput(3.25,1){pq2}{$\state[2]$}
\cnodeput[doubleline=true](6.25,1){pq3}{$\state[3]$}

\ncline{->}{pq1}{pq2}
\goto[\fork{$\atfa$}]{\clock{a}{}{T=0}}
\ncline{->}{pq2}{pq3}
\goto[\join{$\atfa$};\fork{$\btfa$}]{\clock{b}{}{}}


\ncloop[angleB=180,loopsize=1.6]{->}{pq3}{pq1}
\dgoto[\join{$\btfa$}]{\clock{c}{T<25}{}}

\rput(7.5,2){\rnode{A}{$\atfa$:}}
\cnodeput[linestyle=dashed](8.5,2){aq1}{$\bstate[1]$}
\cnodeput(11.5,2){aq2}{$\bstate[2]$}
\cnodeput[doubleline=true](14.5,2){aq3}{$\bstate[3]$}

\ncline{->}{aq1}{aq2}
\goto{\clock{0}{}{\btimer=0}}
\ncline{->}{aq2}{aq3}
\goto{\clock{1}{\btimer<10}{}}
\rput(7.5,1){\rnode{B}{$\btfa$:}}
\cnodeput[linestyle=dashed](8.5,1){bq1}{$\bstate[1]$}
\cnodeput(11.5,1){bq2}{$\bstate[2]$}
\cnodeput[doubleline=true](14.5,1){bq3}{$\bstate[3]$}

\ncline{->}{bq1}{bq2}
\goto{\clock{0}{}{\ctimer=0}}
\ncline{->}{bq2}{bq3}
\goto{\clock{1}{\ctimer<20}{}}
\end{pspicture}
\end{center}
\caption{An inconsistent parallel timing system $\apts[2]$. }
\label{fig:notcons}
\end{figure}

%% file: thm_cons.tex
\begin{example}
$\apts[1]$ is consistent.
\begin{proof}
$\tbaparent'$, the result of flattening $\apts[1]$, is shown below, with $\atimer[\atfa]$ being the
derived timer corresponding to $\atfa$:

\input{thm_cons_pic}

Furthermore, $\uses{{\tbaparent'}}=\set{
	\pair{
		\pair{
			\state[1]
		}{
			\state[2]
		}
	}{
		\pair{
			\state[2]
		}{
			\state[1]
		}
}}$, and by observation, all paths through $\tbaparent'$ beginning with the edge 
$\pair{\state[1]}{\state[2]}$ and ending with
the edge $\pair{\state[1]}{\state[2]}$ take the form $\state[1](\state[2])^*\state[1]$. All such
paths $\apath$ satisfy inequality~\ref{eq:cons}, 
and thus by definition, $\apts[1]$ is consistent.
\end{proof}
\end{example}

%% file: thm_cons_pic.tex
\begin{center}
\begin{pspicture}(0,0)(3,2.5)

\cnodeput[linestyle=dashed](0,0.5){pq1}{$\state[1]$}
\cnodeput[doubleline=true](3,0.5){pq2}{$\state[2]$}

\ncline{->}{pq1}{pq2}
\goto[{$\atimer[\atfa]=0$}]{\clock{a}{}{T=0}}

\ncarc[arcangleA=135, arcangleB=135, ncurv=3]{->}{pq2}{pq2}
\aput{:U}{\clock{b}{}{}}

\ncloop[angleB=180,loopsize=1.6]{->}{pq2}{pq1}
\dgoto[{$\timer{\atimer[\atfa]<10}$}]{\clock{c}{T<50}{}}

\end{pspicture}
\end{center}

%% file: thm_incons.tex
\begin{example}
$\apts[3]$ is not consistent.
\begin{proof}
$\tbaparent'$, the result of flattening $\apts[3]$, is shown in Figure~\ref{fig:s3-flat}.
Furthermore, \[\uses{{\tbaparent'}}=\set{
	\pair{
		\pair{
			\state[1]
		}{
			\state[2]
		}
	}{
		\pair{
			\state[3]
		}{
			\state[1]
		}
}
,
	\pair{
		\pair{
			\state[2]
		}{
			\state[3]
		}
	}{
		\pair{
			\state[3]
		}{
			\state[1]
		}
}}
\]
There are thus two paths against which we need to test inequality~\ref{eq:cons}:
$\path{\state[1]\state[2]\state[3]\state[1]}$, and 
$\path{\state[2]\state[3]\state[1]}$; the first of these fails the test: 
$\delayTBA[\tbaparent']{\state[1]\state[2]\state[3]\state[1]} = 25$,
and
$\delayTBA[\tbaparent]{\state[1]\state[2]\state[3]\state[1]} = 24$.

\end{proof}
\end{example}

%% file: cases.tex
\section{Case Study: Matrix Multiplication}
\label{sec:cases}
We now turn our attention to a practical application of the concepts discussed so far. Namely,
we demonstrate the use of the formal validation concepts on a simple parallel, 
MPI-style \cite{mvapich,mpi} matrix multiplication kernel,
extracted from the larger power-grid analysis application described in \cite{rtss-wip,hui-sc}.
Our kernel implements a variant of Fox's algorithm for matrix multiplication \cite{FoxMM}. For simplicity,
we assume square matrices, and that the number of columns, rows, and processors
are all perfect squares. The algorithm distributes the
task of multiplying two matrices amongst all processors in the system. 

We give a simple distributed algorithm for matrix multiplication, and
a consistent parallel timing system for that algorithm. We conclude the section with empirical
results--- timing measurements taken on a small, four-node real-time cluster, 
each node consisting of dual quad-core 2.66Ghz Xeon X5660 processors running 
the Xenomai RTOS with 48GB RAM.
The timing measurements of the PTS,
along with the usual restrictions associated with real-time computation
(e.g. no virtual memory or paging, process scheduling, ensuring minimal variance in execution timings,
etc.),
are bounded by virtue of Xenomai's real-time process scheduler.
The result is a matrix multiplication kernel which provably runs in under 9 ms per cycle
for $128\times 128$ double-precision matrices.
We emphasize
that we are not claiming the speed of the operation to be a groundbreaking result--- obviously, 
this is a relatively small matrix size, but was so chosen as this is the order of the size 
required by our targeted application kernel. 
Rather, we give these numbers, as well as the PTS,
to illustrate the \emph{process} by which we analyze the temporal interactions between processes,
thus showing this delay to be a provable upper bound.

\subsection{Algorithm}
\begin{algorithm}
{
\begin{tabular}{lll}
$p$&:&Number of processors\\
$N$&:&Rank of matrices
\end{tabular}
}
\caption{MatrixMultiply: Compute $\cmat = \amat \times \bmat$}
\label{alg:mm}
\begin{algorithmic}[1]
\STATE $q \gets \sqrt{p}$ 
\WHILE{\TRUE} \label{algline:mm:while}
\STATE $dest\gets 1$ \label{algline:mm:dest}
\IF[Master process]{self $== 0$}  \label{algline:mm:master}
	\FOR{$i = 0$ to $q-1$}  \label{algline:mm:masterstart}
		\FOR{$j=0$ to $q-1$}
		\STATE $w \gets i\frac{N}{q}, x\gets (i+1)\frac{N}{q}$ \label{algline:mm:partition1}
		\STATE $y \gets j\frac{N}{q}, z\gets (j+1)\frac{N}{q}$ 
		\STATE $\xs \gets \amat[w:x][0:N]$
		\STATE $\ys \gets \bmat[0:N][y:z]$ \label{algline:mm:partition2}
		\IF[Master already has these chunks]{$i\neq 0$ \AND $j \neq 0$} \label{algline:mm:if}
			\STATE $\send{\xs}{dest}$ \label{algline:mm:send1}
			\STATE $\send{\ys}{dest}$ \label{algline:mm:send2}
			\STATE $dest\gets dest + 1$ \label{algline:mm:incdest}
		\ENDIF
		\ENDFOR
	\ENDFOR \label{algline:mm:masterend}
\ELSE[Child processes] \label{algline:mm:child}
	\STATE $\xs \gets \recv{0}$\label{algline:mm:recv1}
	\STATE $\ys \gets \recv{0}$\label{algline:mm:recv2}
\ENDIF
\STATE $\zs \gets \locmm{\xs}{\ys}$\label{algline:mm:locMM}
\STATE $\wb{\zs}{\cmat}$\label{algline:mm:wb}
\ENDWHILE \label{algline:mm:endwhile}
\end{algorithmic}
\label{algo:mm}
\end{algorithm}

The pseudocode for the algorithm is given in Algorithm~\ref{algo:mm}.
Conceptually, to multiply two $N\times N$ matrices $\amat$ and $\bmat$  using a $p$ processor cluster, each matrix is 
divided into 
segments, which are then distributed in round-robin fashion amongst the processors of the cluster.
Each processor then performs a
local matrix multiplication on its own local submatrices, and the results of these local operations are 
aggregated (reduced) to form the matrix product $\amat \times \bmat$.

\begin{figure}[hlt]
\begin{center}
\begin{pspicture}(0,0.5)(8,3)
\psframe(0,1)(2,3) 
\psframe(3,1)(5,3) 
\psframe(6,1)(8,3) 

\psline(0,2)(2,2)

\psline(4,1)(4,3)

\psline(7,1)(7,3)
\psline(6,2)(8,2)

\rput(1,0.5){$\amat$}
\rput(4,0.5){$\bmat$}
\rput(7,0.5){$\cmat$}
\rput(2.5,2){$\times$}
\rput(5.5,2){$=$}

\rput(1,2.5){$\amat_1$}
\rput(1,1.5){$\amat_2$}

\rput(3.5,2){$\bmat_1$}
\rput(4.5,2){$\bmat_2$}

\rput(6.5,2.5){$\cmat_1$}
\rput(7.5,2.5){$\cmat_2$}

\rput(6.5,1.5){$\cmat_3$}
\rput(7.5,1.5){$\cmat_4$}
\end{pspicture}
\end{center}
\caption{Partitioning and distribution of matrix multiplication by Algorithm~\ref{algo:mm} across a four
processor cluster.  
}
\label{fig:mm-ex}
\end{figure}
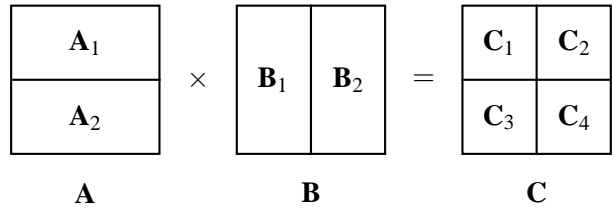

Due to space constraints, we will not describe the partitioning in detail;
Figure~\ref{fig:mm-ex} shows the partitioning and distribution of work by Algorithm~\ref{algo:mm} for a four-processor cluster.
In this figure, $\amat,\bmat,$ and $\cmat$ are all $N \times N$ matrices. $\amat$ is partitioned into
2  sets of $\frac{N}{2}$ rows each, and  $\bmat$ is partitioned into 2 sets of $\frac{N}{2}$ columns each.
The master process, $p_0$, computes the local product $\amat_1\times\bmat_1$, and writes the result to $\cmat_1$.
$p_0$ then sends submatrices $\amat_1$ and $\bmat_2$ to $p_1$, who 
then computes their product, writing the result to $\cmat_2$.
Similarly, 
$p_2$ receives and computes $\cmat_3=\amat_2\times\bmat_1$,  and
$p_3$ receives and computes $\cmat_4=\amat_2\times\bmat_2$. 


The algorithm proceeds as follows: the master process 
executes lines \ref{algline:mm:masterstart}  through \ref{algline:mm:masterend},
which partition $\amat$ and $\bmat$ into submatrices 
(lines \ref{algline:mm:partition1}--\ref{algline:mm:partition2}), 
and send these parts out to the respective child processes (lines
\ref{algline:mm:send1}--\ref{algline:mm:send2}).
Conversely, the child processes execute lines~\ref{algline:mm:recv1}--\ref{algline:mm:recv2}, which 
receive the
submatrices assigned by the master process.
Lines~\ref{algline:mm:locMM}--\ref{algline:mm:wb} are run
by all processes, including the master process (which, in this case, participates
in the task of matrix multiplication as well). Line \ref{algline:mm:locMM}
performs the local operation, line~\ref{algline:mm:wb} writes the local
result to the appropriate location in $\cmat$.
The entire process then repeats indefinitely, as given by the
\textbf{while} loop (lines~\ref{algline:mm:while} and \ref{algline:mm:endwhile}).

%
\subsection{Parallel Timing System}
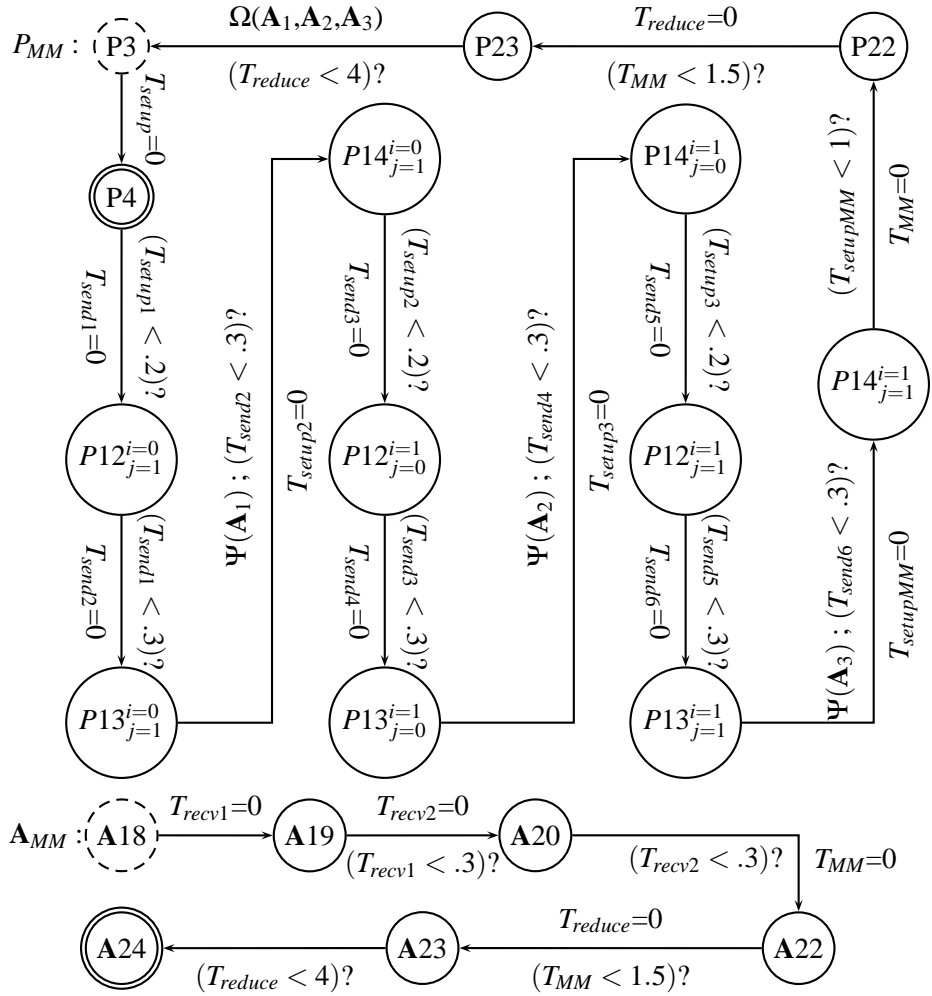
\begin{figure*}
\begin{center}
\input{parentDFA}
\end{center}
\caption{Parallel timing system $\apts[MM]$ for Algorithm~\ref{algo:mm} across a four processor cluster.
Events have been elided for the sake of clarity. Upper bounds on timer constraints correspond to delay measurements
taken over our implementation; times are given in milliseconds.
Minimal variance from these bounds is ensured to the extent
provided by the underlying RTOS.}
\label{pts:mm}
\end{figure*}

Figure \ref{pts:mm} shows a parallel timing system for Algorithm~\ref{algo:mm} across
a four processor cluster, consisting
of the TBA $\tbaparent[MM]$, which models the master process, and a child TFA $\tbachild[MM]$,
modeling instances of the child processes. Specific events have been elided from the diagram
in this case, since events in this case always represent transitions between statements.

\subsubsection{Parent}
States in the parent automaton $\tbaparent$ 
are prefixed with a `P', followed by
the line number as given in Algorithm~\ref{algo:mm}.
For example, $P\ref{algline:mm:dest}$ corresponds to
the state of the parent process as it is executing line \ref{algline:mm:dest}.

Additionally, lines \ref{algline:mm:send1} and \ref{algline:mm:send2} each beget
three separate states--- parameterized on the values of the loop induction
variables $i$ and $j$--- and are labeled accordingly. As is commonly the case
in WCET analysis, unrolling the loop nest in this
fashion is necessary in order to obtain a strict upper bound on the number of
iterations and, consequently, the total execution time, of the loop nest.

$\tbaparent$ forms, in this case, a simple cycle. The cycle starts at state
$\tbaparent\ref{algline:mm:dest}$, and steps sequentially through the steps
(states) of the algorithm.  Namely, the parent process starts at line $\ref{algline:mm:dest}$
(i.e., state $\tbaparent\ref{algline:mm:dest}$),
and proceeds sequentially through lines $\ref{algline:mm:master}$ 
(state $\tbaparent\ref{algline:mm:master}$), and eventually
to line $\ref{algline:mm:send1}$ ($\stij{\tbaparent\ref{algline:mm:send1}}{0}{1}$). The delay between 
the initialization 
(state $\tbaparent\ref{algline:mm:dest}$) and the first send 
($\stij{\tbaparent\ref{algline:mm:send1}}{0}{1}$) is
bounded by a timer, $\atimer[setup1]$ (the idea being that this is the delay incurred
by the time to ``set up" the first send). 
Execution then proceeds to line~$\ref{algline:mm:send2}$ 
($\stij{\tbaparent\ref{algline:mm:send2}}{0}{1}$); the delay along this transition
represents the time to send the first chunk to the respective child process, and is
bounded by timer $\atimer[send1]$. At this point, execution proceeds to 
line~$\ref{algline:mm:incdest}$ ($\stij{\tbaparent\ref{algline:mm:incdest}}{0}{1}$). Along this transition, there are two items to note:
first, the time to process the second send is bounded by  the timer $\atimer[send2]$,
and second, the child process has now been sent the data it needs,
and consequently, $\tbachild[1]$ is forked. Execution proceeds similarly through the
next six states, representing the unwound iterations of the loop nest. 
Child process $\tbachild[2]$ is similarly forked on the transition from $\stij{\tbaparent\ref{algline:mm:send2}}{1}{0}$
to $\stij{\tbaparent\ref{algline:mm:incdest}}{1}{0}$, and
$\tbachild[3]$ on the transition from $\stij{\tbaparent\ref{algline:mm:send2}}{1}{1}$
to $\stij{\tbaparent\ref{algline:mm:incdest}}{1}{1}$.
Execution then proceeds through lines
$\ref{algline:mm:locMM}$ (state $\tbaparent\ref{algline:mm:locMM}$) and $\ref{algline:mm:wb}$
($\tbaparent\ref{algline:mm:wb}$).
The duration of the local 
matrix multiplication operation (line~$\ref{algline:mm:locMM}$) is bounded by the timer
$\atimer[MM]$, and that through the reduce operation (line~$\ref{algline:mm:wb}$)
by the timer $\atimer[reduce]$. Additionally, the transition from   
$\tbaparent\ref{algline:mm:wb}$ back to   $\tbaparent\ref{algline:mm:dest}$ waits
for (joins with) all child processes to complete before proceeding.

\subsubsection{Child}
In this case, the child processes are modeled by the TFA $\tbachild$.
Nomenclature is analogous to that of $\tbaparent$: states in $\tbachild$ are
prefixed with an $\tbachild$, followed by the corresponding line number  from 
Algorithm~\ref{algo:mm}.

The child process starts at line $\ref{algline:mm:child}$ (state $\tbachild\ref{algline:mm:child}$).
The process then proceeds to receive the first block of data (line $\ref{algline:mm:recv1}$,
state $\tbachild\ref{algline:mm:recv1}$). The time to process the receive is bounded
by timer $\atimer[recv1]$. Execution proceeds to receive the second block of data
 (line $\ref{algline:mm:recv2}$, state $\tbachild\ref{algline:mm:recv2}$). The time
 to process this second receive is bounded by $\atimer[recv2]$.  Execution proceeds  next
 to the local matrix multiplication 
 (line $\ref{algline:mm:locMM}$, state $\tbachild\ref{algline:mm:locMM}$); the time
 spent on this operation is bounded by timer $\atimer[locMM]$.
 Finally, execution proceeds to the data writeback 
 (line $\ref{algline:mm:wb}$, state $\tbachild\ref{algline:mm:wb}$); the time
 spent on this operation is bounded by timer $\atimer[reduce]$.

\begin{theorem}
$\apts[MM]$ is consistent.
\begin{proof}
Let $\flatten{\apts[MM]}=\pair{\ireln'}{\creln'}$,
with $\tbaparent[MM]'=\tba{\Aalpha}{\astates}{\state[0]}{\accepts}{\aclks}{\tbareln}{\ireln'}{\creln'}$. By definition, 
$\uses{\apts[MM]} = \set{
		\pair{\aedge[1]}{\aedge[4]},
		\pair{\aedge[2]}{\aedge[4]},
		\pair{\aedge[3]}{\aedge[4]}}$, where
\begin{align*}
	\aedge[1] &= \pair{\stij{\tbaparent{\ref{algline:mm:send2}}}{0}{1}}{\stij{\tbaparent{\ref{algline:mm:incdest}}}{0}{1}} &
	\aedge[2] &= \pair{\stij{\tbaparent{\ref{algline:mm:send2}}}{1}{0}}{\stij{\tbaparent{\ref{algline:mm:incdest}}}{1}{0}}\\
	\aedge[3] &= \pair{\stij{\tbaparent{\ref{algline:mm:send2}}}{1}{1}}{\stij{\tbaparent{\ref{algline:mm:incdest}}}{1}{1}} &
	\aedge[4] &= \pair{\tbaparent{\ref{algline:mm:wb}}}{\tbaparent{\ref{algline:mm:dest}}}
\end{align*}
By observation, there are three paths which we must consider:
\begin{align*}
\apath[1] &= \path{
		   \stij{\tbaparent{\ref{algline:mm:send2}}}{0}{1}
		   \stij{\tbaparent{\ref{algline:mm:incdest}}}{0}{1}...
		   \tbaparent\ref{algline:mm:wb}
		   \tbaparent\ref{algline:mm:dest}
}\\
\apath[2] &= \path{
		   \stij{\tbaparent{\ref{algline:mm:send2}}}{1}{0}
		   \stij{\tbaparent{\ref{algline:mm:incdest}}}{1}{0}...
		   \tbaparent\ref{algline:mm:wb}
		   \tbaparent\ref{algline:mm:dest}
}\\
\apath[3] &= \path{
		   \stij{\tbaparent{\ref{algline:mm:send2}}}{1}{1}
		   \stij{\tbaparent{\ref{algline:mm:incdest}}}{1}{1}...
		   \tbaparent\ref{algline:mm:wb}
		   \tbaparent\ref{algline:mm:dest}
}
\end{align*}

The rest of the proof follows by enumeration:
\begin{align*}
	(\delayTBA[{\tbaparent[MM]'}]{\apath[1]} &= 6.1)  <  (\delayTBA[{\tbaparent[MM]}]{\apath[1]} = 8.1)     \\
	(\delayTBA[{\tbaparent[MM]'}]{\apath[2]} &= 6.1)  <  (\delayTBA[{\tbaparent[MM]}]{\apath[2]} = 7.3)     \\
	(\delayTBA[{\tbaparent[MM]'}]{\apath[3]} &= 6.1)  <  (\delayTBA[{\tbaparent[MM]}]{\apath[3]} = 6.5) 
\end{align*}

\end{proof}
\label{thm:mmcons}
\end{theorem}

Finally, we note that the worst case delay along one iteration of the algorithm
is 8.9 ms. This follows from the observation that the parent automaton $\tbaparent$
takes the form of a simple cycle with no unbound segments
(i.e., subpaths which are not constrained by any timer). Specifically,
$\tbaparent$ consists of consecutive pairs of segments, each constrained by pairs of timers.
Consequently, we 
can derive an upper bound for a single iteration of the algorithm by summing
the bounds of all of the timers, yielding the specified upper bound.
Combined with Theorem~\ref{thm:mmcons}, which ensures that the timing of the child processes does 
not invalidate this bound, we are left with a cyclic, parallel, time-bounded matrix multiplication 
kernel.

%% file: parentDFA.tex
\begin{pspicture}(11,12.5)
\rput(0,12.5){$\tbaparent[MM]:$}
\cnodeput[linestyle=dashed](1,12.5){Pdest}{P\ref{algline:mm:dest}}
\cnodeput[doubleline=true](1,10.5){Pmaster}{P\ref{algline:mm:master}}

\ncline{->}{Pdest}{Pmaster}
\goto{\atimer[setup]=0}

\cnodeput(1,7){send1-01}{\stij{\tbaparent\ref{algline:mm:send1}}{0}{1}}

\ncline{->}{Pmaster}{send1-01}
\goto[{\atimer[send1]=0}]{\timer{\atimer[setup1]<.2}}
\cnodeput(1,3.5){send2-01}{\stij{\tbaparent\ref{algline:mm:send2}}{0}{1}}

\cnodeput(4.5,11){incdest01}{\stij{\tbaparent\ref{algline:mm:incdest}}{0}{1}}
\cnodeput(4.5,7){send1-10}{\stij{\tbaparent\ref{algline:mm:send1}}{1}{0}}
\cnodeput(4.5,3.5){send2-10}{\stij{\tbaparent\ref{algline:mm:send2}}{1}{0}}

\cnodeput(8.5,11){incdest10}{\tbaparent\stij{\ref{algline:mm:incdest}}{1}{0}}
\cnodeput(8.5,7){send1-11}{\stij{\tbaparent\ref{algline:mm:send1}}{1}{1}}
\cnodeput(8.5,3.5){send2-11}{\stij{\tbaparent\ref{algline:mm:send2}}{1}{1}}

\ncline{->}{send1-11}{send2-11}
\goto[{\atimer[send6]=0}]{\timer{\atimer[{send5}]<.3}}


\cnodeput(11,8){incdest11}{\stij{\tbaparent\ref{algline:mm:incdest}}{1}{1}}
\cnodeput(11,12.5){locMM}{\tbaparent\ref{algline:mm:locMM}}
\cnodeput(6,12.5){wb}{\tbaparent\ref{algline:mm:wb}}

\ncline{->}{11-11}{incdest11}
\ncangle[angleA=0,angleB=270]{->}{send2-11}{incdest11}
\goto[{\atimer[setupMM]=0}]{\clock{}{\atimer[{send6}]<.3}{\fork{\tbachild[3]}}}
\ncline{->}{incdest11}{locMM}
\goto[{\atimer[MM]=0}]{\timer{\atimer[setupMM]<1}}

\ncline{->}{send1-01}{send2-01}
\goto[{\atimer[send2]=0}]{\timer{\atimer[{send1}]<.3}}

\ncangle[angleA=0,angleB=180,arm=.75cm]{->}{send2-01}{incdest01}
\goto[{\atimer[setup2]=0}]{\clock{}{\atimer[{send2}]<.3}{\fork{\tbachild[1]}}}

\ncline{->}{P2-2}{P2-3}
\ncline{->}{incdest01}{send1-10}
\goto[{\atimer[send3]=0}]{\timer{\atimer[{setup2}]<.2}}

\ncline{->}{send1-10}{send2-10}
\goto[{\atimer[send4]=0}]{\timer{\atimer[{send3}]<.3}}
\ncangle[angleA=0,angleB=180,arm=.75cm]{->}{send2-10}{incdest10}
\goto[{\atimer[setup3]=0}]{\clock{}{\atimer[{send4}]<.3}{\fork{\tbachild[2]}}}
\ncline{->}{locMM}{wb}
\dgoto [\timer{\atimer[{MM}]<1.5}] {{\atimer[reduce]=0}}
\ncline{->}{incdest10}{send1-11}
\goto[{\atimer[send5]=0}]{\timer{\atimer[{setup3}]<.2}}

\ncline{->}{wb}{Pdest}
\dgoto[\timer{\atimer[reduce]<4}]{\join{\tbachild[1],\tbachild[2],\tbachild[3]}}

\rput(0,2){$\tbachild[MM]:$}
\cnodeput[linestyle=dashed](1,2){Cstart}{\tbachild\ref{algline:mm:child}}
\cnodeput(3.5,2){Crecv1}{\tbachild\ref{algline:mm:recv1}}
\cnodeput(6.5,2){Crecv2}{\tbachild\ref{algline:mm:recv2}}
\cnodeput(10,0.5){Cmm}{\tbachild\ref{algline:mm:locMM}}
\cnodeput(5,0.5){Cwb}{\tbachild\ref{algline:mm:wb}}
\cnodeput[doubleline=true](1,0.5){Cend}{\tbachild\ref{algline:mm:endwhile}}
\ncline{->}{Cstart}{Crecv1}
\goto{\atimer[recv1]=0}
\ncline{->}{Crecv1}{Crecv2}
\goto[\timer{\atimer[recv1]<.3}]{\atimer[recv2]=0}
\ncangle[angleA=0,angleB=90]{->}{Crecv2}{Cmm}
\cgoto[\timer{\atimer[recv2]<.3}]{\atimer[MM]=0}
\ncline{->}{Cmm}{Cwb}
\dgoto[\timer{\atimer[MM]<1.5}]{\atimer[reduce]=0}

\ncline{->}{Cwb}{Cend}
\dgoto[\timer{\atimer[reduce]<4}]{}

\end{pspicture}

%% file: remarks.tex
\section{Concluding Remarks}
\label{sec:concl}
We conclude with a few closing remarks. We have presented a formal system for modeling the temporal
properties of a restricted class of real-time parallel systems, with a simple example of an 
application kernel. 
As is usually the case with real-time systems, 
loops need to be unrolled, bounding the number of iterations, in order 
to obtain an upper bound on the total execution time of the loop. 
Algorithm~\ref{algo:mm} (intentionally) distills to a relatively simple PTS, due to the basic structure
of the control flow graph of both the parent and child processes; more complex 
examples are of obvious interest for future work.
Similarly, the model in Figure~\ref{pts:mm} in our case was derived manually--- in this case,
a relatively simple task. More complex examples can certainly prove to be more of a challenge,
and automated tools for this task are desirable. One possible approach for such automation
would be compiler-driven, whereby users could specify to the compiler (via {\tt \#pragma}s, 
for instance), events of interest, and the compiler could proceed to output the appropriate
annotated control flow graph. 

We assume timing behavior is consistent across all child processes, although if there were to
be significant variance across child processes (e.g. heterogeneous or NUMA architectures) we 
could account for such behavior using different child TFAs.

Additionally, we have laid out several interesting open questions 
which arise out of the analysis of our relatively straightforward formulation: 
what is the complexity of computing the worst case delay along a single path of a TFA (TBA), and
through a TFA (TBA) in general?  Up to this point, we have only considered conjunctions of
maximum constraints; how does this change in the presence of a more generalized constraint
syntax (c.f. \cite{tba})?

We have largely been working with the SPMD execution model paradigmatic
of many MPI-type programs.  It would be interesting to investigate temporal models for 
other parallel models (e.g. OpenMP) as well.
Lastly, our application kernel distills to a 
relatively simple set of automata. More complex examples are certainly of interest, and
are on the horizon for future work.